# SmartOil: Blockchain and Smart Contract-based Oil Supply Chain Management

AKM Bahalul Haque
Lecturer, Dept. of Electrical and Computer Engineering,
North South University
Email:Bahalul.haque@northsouth.edu

Md. Rifat Hasan
Dept. of Electrical and Computer Engineering,
North South University
rifat.hasan03@northsouth.edu

Md. Oahiduzzaman Mondol Zihad
Dept. of Electrical and Computer Engineering,
North South University
mondol.zihad@northsouth.edu

**Corresponding Author:**

AKM Bahalul Haque
Lecturer, Dept. of Electrical and Computer Engineering,
North South University
Email:Bahalul.haque@northsouth.edu

**Postal Address:** Room: SAC 11100, Dept. Electrical and Computer Engineering, North South University, Dhaka, Bangladesh -1229

Funding: None

Conflict of Interest statement:The authors declare no conflict of interest

Permission to reproduce materials from other sources: None

Data Availability statement: Data/Code is uploaded while submitting the manuscript

## Abstract

The traditional oil supply chain suffers from various shortcomings regarding crude oil extraction, processing, distribution, environmental pollution and traceability. It offers only forward flow of products with almost no security and tracking process. In time, the system will lag behind due to the limitations in quality inspection, fraudulent information and monopolistic behavior of supply chain entities. Inclusion of counterfeiting products and opaqueness of the system urge renovation in this sector. The recent evolution of Industry 4.0 leads to the alternation in the supply chain introducing the smart supply chain. Technological advancement can now reshape the infrastructure of the supply chain for the future. In this paper, we suggest a conceptual framework utilizing Blockchain and Smart Contract to monitor the overall oil supply chain. Blockchain is a ground-breaking technology to monitor and support the security building of a decentralized type supply chain over a peer-to-peer network. The use of Internet of Things (IoT), especially sensors, opens a broader window to track the global supply chain in real-time. We construct a methodology to support reverse traceabilty for each participant of the supply chain. The functions and characteristics of Blockchain and Smart Contract are defined. Implementation of Smart Contracts has also been shown with detailed analysis. We further describe the challenges of implementing such a system and validate our framework's adaptability in the real world. The paper concludes with future research scope to mitigate the restrictions of data management and maintenance with advanced working prototypes and agile systems achieving greater traceability and transparency.

**Keywords:** Smart Supply Chain Management, Blockchain, Smart Contract, IoT devices, Oil Supply Chain.

## 1. Introduction

Supply Chain Management (SCM) in the oil industry holds great importance for its expansion and global market. The value of oil has a significant impact on the world economy. In general, SCM can be coined as the flow of ingredients from production to consumer. It includes several phases from supplying raw materials to end customer including manufacturer, distributor, retailer, etc. For the oil supply chain, these entities are an industry or a company. Moreover, it is a global process as the crude oil is extracted from one place, processed and refined at another, and distributed worldwide. The traditional oil supply chain serves general purposes but cannot comply fully. There are certain limitations in tracking, monitoring, and giving the end customer power to reverse track. It generally goes with the forward flows i.e., the raw materials' flow to final products (Li & Olorunniwo, 2008; Prahinski & Kocabasoglu, 2006). However, it is equally important to support the reverse flow of product information for any customer.

Blockchain and smart contracts, along with IoT devices, can change the traditional supply chain for oil. Blockchain can support the supply chain with the benefits of transparency and immutability (Apte & Petrovsky, 2016). Blockchain helps the supply chain to be modernized by providing a secured system for recording data and implement and run coded scripts or applications called smart contracts (Alqahtani et al., 2020). With the power of smart contracts and IoT devices, the oil supply chain can have the ability of provenance and security. We discuss these issues and try to configure a possible solution.

Our approach in this paper is to model a conceptual framework for the oil supply chain. Our goal is to utilize the power of blockchain and smart contracts to conduct necessary tasks to ensure traceability. The IoT devices further empower the system to update and check the smart contract. It allows every customer to check back the product information in case of oil. Most importantly, our framework establishes a complete monitoring system in oil supply chain management.

We explain our work in the rest of the paper with the following sections. In section 2, we provide a brief discussion about the smart supply chain and the use of blockchain along with the smart contracts. Section 3 gives an insight into the recent works on the oil supply chain. In section 3, we point some significant issues that the existing system has. Section 5 and 6 explain the framework in detail with detailed applications. Later, we show an implementation of the proposed smart contract and analysis with advantages and challenges in sections 7 and 8. Finally, section 7 discusses the future work and conclusion.

## 2. Background Overview of Smart Supply Chain

The shortcomings of the traditional supply chains advance the world to modernize the supply chain. On the other hand, Industry 4.0 expands the area of application for the smart supply chain. The innovation of the latest technologies also broadens the path. Blockchain and smart contracts are the most prominent among them.

### 2.1 Blockchain and Smart Contract

The concept of Blockchain comes from the idea of an immutable and decentralized system. It is a distributed ledger technology (DLT) for a trustless environment(Prashanth Joshi et al., 2018) .

Blockchain is a chain of blocks that can store records of transactions in an untrusted environment. Blockchain does the trust in such an environment by validating each block in the mining process. In Blockchain all the members are treated like a server and they are treated equally. They can share data directly peer to peer (P2P). In this system user doesn't need any medium or centralized server to send or receive. Blockchain is a reliable space to store and share data . Blockchain innovation is a mix of different calculations and methods like cryptography and decentralized system. Blockchain can be characterized as the incorporated framework of multifield applications like healthcare, financial sector, IoT, smart infrastructures, etc. Accessibility of information is consistent with better quality in these blockchain frameworks. Blockchain has some unique attributes for example data inside a blockchain is immutable, tamper-proof, built on a decentralized network, data is hashed and cryptographically secured, etc. Fundamentally, there are three types of blockchain namely public or permissionless, private of permissioned, and consortium blockchain. Each of them has specific characteristics through the uniqueness lies in the geographical region of the network (Bodkhe et al., 2020; Compagno et al., 2020).

On the other hand, a smart contract takes blockchain one step further by imparting the ability to generate code inside the blockchain. Smart contracts are the code of agreement or rules set by the participants without a third party. The introduction of solidity, a Turing complete and general purposed language, based on Ethereum Virtual Environment (EVM) makes it easy to deploy with blockchain technology(Gupta et al., 2020) . When the smart contract address receives a request, it executes the functions written inside or calls other smart contracts if programmed that way. Eventually, it can keep track of any violation in the process. Smart contracts are not actually a part of Blockchain. In fact, it is entirely a different module that uses blockchains to automate transactions between users and machines. The big advantage of implementing blockchain technology with smart contracts is to create the "peer-to-peer market". Smart Contracts can reduce risks, cut down administration and service costs and improve the efficiency of business processes.

### 2.2 Smart Supply chain

To solve the issues of traditional supply chain such as: improper tracking, monitoring, maintenance, security, and environmental pollution, safety risks etc. Smart Supply Chain is being introduced. With the globalization of supply chain, latest technologies like RFID, QR code etc. are already in use. There are technologies that are reshaping transportation, processing, and procurement. To comply with the current and future needs, smart supply chain adopts smart infrastructure, smart machine, IoT devices. They help the system in smart decision-making and intelligent response to consumer . There are a lot of works to improve the supply chain in various fields based on technologies like blockchain, smart contract, IoT devices and so on (Pranto et al., 2021; Tayal et al., 2021) .

## 3. Literature Review

In this section, we discuss some works on smart supply chain proposing framework or techniques using blockchain and smart technologies. There are multiple works on supply chain management based on other technologies like IoT devices and sensor-based technologies. However, blockchain brings a revolution in the process of resolving several critical challenges.

Edgar et al. [11] discussed the usefulness of reverse logistics i.e., backward product flows and forward logistics in the palm oil supply chain. The authors proposed a closed-loop framework keeping the “green operations” in mind. It aims to ensure environmental sustainability in both forward and reverse flow. The paper also shows some mathematical modeling and its implementation resulting in a positive output over the supply chain using statistical tools. But the industrial level implementation and data security and management are not shown. Kshetri [12] discussed the potential of blockchain in resolving the key objectives of the supply chain e.g., speed, cost, risk, quality products etc. It also explored the connection of IoT devices in a blockchain-based solution for the supply chain. The application in some well-known use cases of the supply chain worldwide like Alibaba, Maersk etc. has been discussed thoroughly. But it lacks a complete framework for the global oil supply chain.

Nehai et al. [13] introduce a novel approach to validate smart contracts. It follows a model-checking method. The model explores several rules to convert smart contract for model checking. The method works in three layers to illustrate the nature, logic and execution of the smart contract. It uses NuSMV tool to write and execute the coded script for smart contracts. It is good approach to validate smart contracts but the interconnection of IoT devices along with blockchain management still remains unsolved. Arena et al. [14] propose a traceability monitoring architecture named BRUCHETTA. It explains the system in a use case – Extra Virgin Olive Oil (EVOO). The proposed method utilizes IoT devices to enable a consumer to access the product information at each phase of oil supply chain. Dividing into six processes, Hyperledger Fabric Blockchain controls the overall supply chain.

Salah et al. [15] suggest an Ethereum blockchain for managing the tracking system in the agricultural system. It designs functions of smart contract to track soybean. The solution uses blockchain to store the information of product for transparency with the help of a decentralized file system. Mavridou et al. [16] introduce the VeriSolid framework to verify smart contracts. It extracts the nature or behavior of the code of smart contract and takes a correct-by-design approach for validation. The framework uses model checking tools like NuSMV. It can directly interact with the Ethereum platform verifying the whole system.

Gupta et al. [17] illustrate the security and privacy of artificial intelligence-based smart contract used in a reatl market environment. Artificial intelligence can be a pioneering tool to ensure the security of smart contracts increasing the efficiency. A comprehensive use case of retail marketing is presented upon discussing the vulnerabilities of smart contracts in a blockchain environment. Vishnubhotla et al. [18] conduct a risk management survey on a case company with blockchain. It prioritizes the risk factors faced in departments like finance, operations, human resources etc. by interviewing the top practitioners in the oil industry. The results show that blockchain adoption can be a potential solution for reducing the risks in supply chain management.

Tayal et al. [19] introduce a novel 3-stage methodology to integrate blockchain in food supply chain. The 3 stages are: principal component analysis, total iterative structureal modeling, and matrice d'impacts croises multiplication appliquee a un classement analysis. Following these steps, nine critical success factors are found that can provide a better scenario increasing efficiency in policy-making for blockchain-integrated food supply chain. Odachi [20] illustrate the obstacles in supply chain management in a Sub-Saharan African oil and gas company.

The paper describes the idea of blockchain Hyperledger to secure and manage the participants' data and automatize the procurement process of the oil supply chain. It discusses the upstream flow of the supply chain in the transaction and smart contracts. Although it lacks the common people usage and global application, the idea shows the efficacy of blockchain technology in supply chain management.

## 4. Issues with Existing System

The concept of supply chain is not a very new one. But it has changed with time. The latest innovation of 21$^{st}$ century and the development of IoT systems modernize present-day supply chain management. The supply chain for crude oil is a global process and there are still significant issues in several aspects.

- Trust: Any transaction needs to maintain a trusted atmosphere. But in the trustless environment of present-day business, there is almost no scope to ensure trust entirely and monitor it simultaneously.
- Third-Party Dilemma: The intermediates in a supply chain can hold much power in the transaction process. They can disrupt the market value without informing the actual participants of the supply chain and extract profits that eventually affect the end customer
- End-to-End Monitor: In the current system, it is impossible to monitor the status of oil from the extraction process to consumption. The IoT-enabled system can store information, but there needs much paperwork for mutual agreements and product shipment.
- Encrypted System: Modern-day supply chain lacks an encrypted system to store its crucial information like workers' data, agreements among companies, local jurisdictions etc. This can lead to cyber attacks that can hamper the international oil market.
- Human Error: There are many phases where documents need to be filled up manually. The monitoring process is also based on humans in a global transaction. It is more prone to human error that can increase price without any reason and very little chance actually to track the error.
- Different Jurisdiction: In a global supply chain, there is no uniform jurisdiction. So, the product has to pass through different laws in different countries. There are authorities to maintain it but it leads to a centralized system.

So, there are many cases seen where the supply chain cost makes a significant impact on the price hike of global oil due to its existing system issues.

## 5. Conceptual Framework For Oil Supply Chain Management

This section will discuss our framework – SmartOil in detail. This is a conceptual framework for further use in several sectors of smart supply chain.

### 5.1 Overview of the Proposed Framework

Our proposed framework uses private blockchain architecture. To store the smart contracts, we use two types of blockchain: permissioned and consortium. It ensures only the trusted entities to enter the network providing end-to-end visibility. Again, we use private blockchain to store the data collected by IoT devices. It gives each participant in the network to keep track their own product. Moreover, participants can have the power to select the information to show and hide otherwise.

Iot devices are responsible for collecting data. This data will be used in various phases of supplychain i.e smart contracts shall use this data to execute different tasks. This framework is proposed specially for traceability, security, temperature and pressure monitoring for oil transportation. IoT devices like GPS and RFID can serve the purpose of traceability and security. There are several IoT sensor that can wirelessly interact to monitor temperature, pressure, humidity etc. Some of them are already being used in multiple industries. We can use these sensor devices to build a wireless sensor network to collect real-time information of oil transportation. The collected information will be updated to a private blockchain and constantly checked with the conditions of smart contract. These functions of different IoT devices can be controlled together with microcontrollers like Arduino or Raspberry Pi. If there is a violation of smart contract conditions, the buyer and seller can instantly receive the violation message.

Smart contract is deployed in the blockchain so that the contract can be temperproof. In the next subection we have mentioned about the actors of the system. Smart contract shall be deployed between each transaction of the two actors. The seller will generate the smart contract explaining the agreements with the buyer. Smart contract holds the necessary conditions to monitor the product. General information like buyer’s address, seller’s address, product information and history, required temperature, pressure, quantity, cost etc. are coded into smart contract. The seller also needs to provide a public and private key to authenticate the receiver. As soon as buyer confirms the accept message with the private key, smart contract will execute and transaction will also be made. Buyer and seller can also use a passphrase to confirm the transaction.

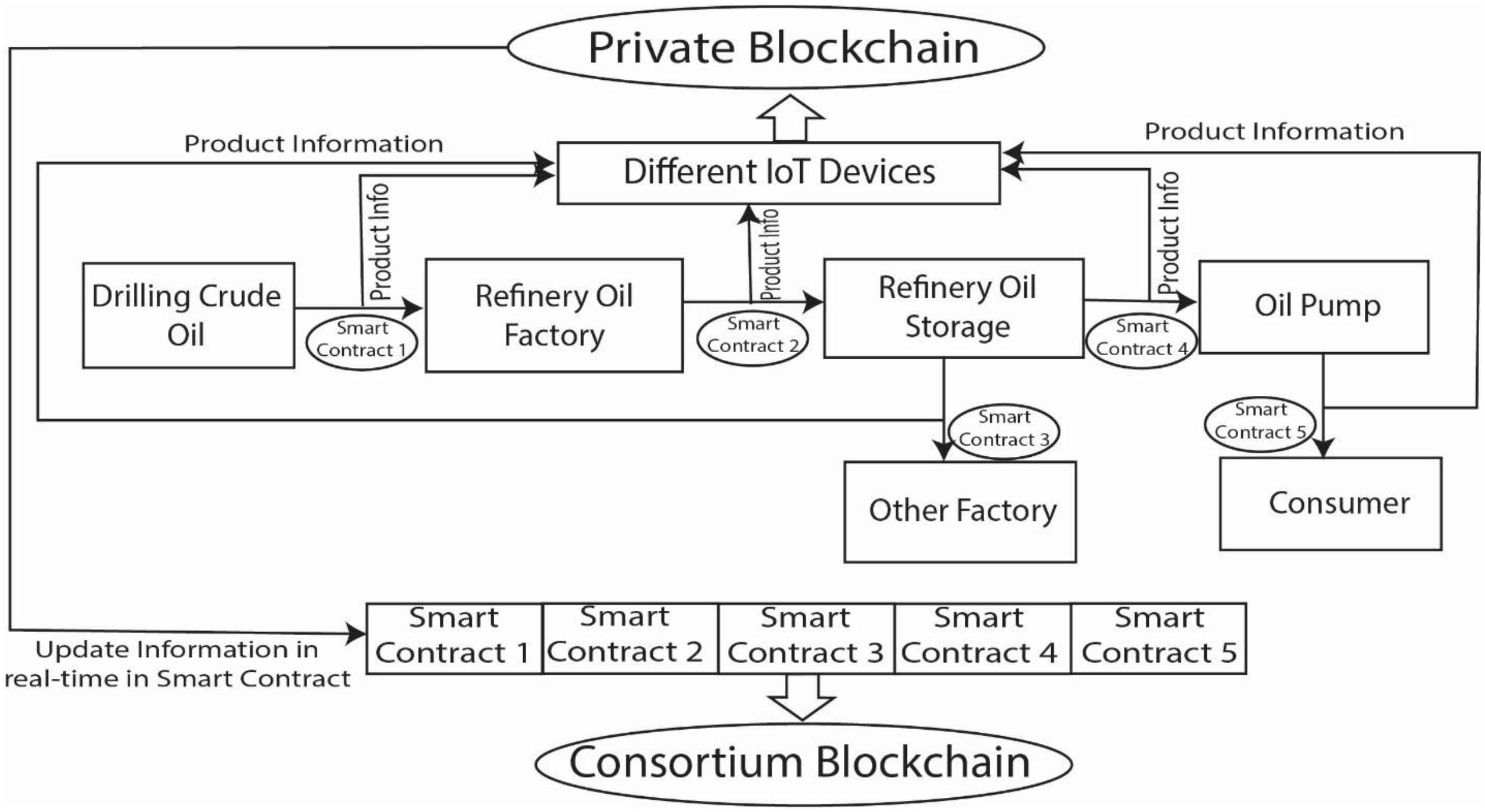

**Figure 1:** Proposed framework for Smart Oil Supply Chain Management

Figure 1 depicts the complete framework and interaction of Smart Contract with Blockchain and IoT devices.
Smart contract will use the private blockchain for real-time data and update. Since blockchain uses cryptographic algorithms and hash functions by default, the data inside blockchain is already very secure and well protected. So we are not implementing any other encryption technique other than the default one used in blockchain.

The function of the whole system is described step-by-step as follows:

- First, the drilling company will initiate a smart contract with refinery factory mentioning the conditions, quantity and product-related information (Smart Contract 1). Upon approval from both party, the shipping will start and smart contract for tracking (collects data by IoT devices) will be initiated by the drilling company for both parties' to check.
- Both the refinery company and drilling company can check if there is any violation of quality from the smart contract for tracking.

- Then, refinery factory will initiate its own smart contract mentioning the product information with storage company (Smart Contract 2). Similarly, both party can the violation in shipping from tracking smart contract.
- Stroage company can initiate two different smart contracts (Smart Contract 3 and 4) with other factories i.e., textile or mills and oil pumps. Different smart contracts for tracking will also be initiated.
- Finally, oil pump will initiate product-information smart contract for consumer and consumer check if any violation is occurred previously.
- All the smart contracts will hold the information of its previous smart contract for violation in quality from tracking smart contract (Consortium Blockchain). But the product-information smart contract will only be accessed by the permissioned participants (Private Blcockhain).

In this paper, we introduce two generalized smart contracts for product information and violation tracking. Participants can modify them as per needs.

**5.2 Actors of the Proposed Framework**

We consider six actors in the oil supply chain – drilling companies, refinery factories, refined oil storage, oil pump, other factories and common consumer.

i. *Drilling Companies:* Drilling companies are responsible for exploring the crude oil from underground. They are the primary source of oil. They will only sell the crude oil to refinery companies. There can be multiple types and qualities of crude oil. The information related to the oil must be apprised to the refinery factories. Drilling companies are like supplier to the supply chain.

ii. *Refinery Factories:* Refinery factories will refine the crude oil. They are responsible to extract different types of oil e.g., food oil, fuel for transports etc. Refinery factories will buy crude oil from drilling companies and sell the refined oil to the refinery storage companies. If the refinery factories hold its own storage company, then they may sell directly to other companies or oil pumps. Refinery companies can be considered as the manufacturer of supply chain.

iii. *Refinery Oil Storage:* The main purpose of Oil Storage company is to store various types of oil and provide them to different pump station or other factories. They buy from refinery factories. They are the logistic operators or distributors of the system.

iv. *Oil Pump:* Oil pump will buy product from refinery storage and sell it to other customer. They only export or buy oil and sell it. There is no manufacturing or processing stage for oil pump. They can be considered as the retailers.
v. *Other Factories:* They will only buy oil from refinery storage or oil pump for various purpose like transportation, production etc.
vi. *Common Consumer:* They are the end users are the common consumers of the system

## 6. Application Scenario

In this section, we will detail the application of our framework in oil supply chain. We divide the applications in three general categories: traceability and tracking procedure, temperature monitoring, and global and local supply chain.

### 6.1 Traceability and Tracking Procedure:

Our framework uses IoT devices and sensors to conduct tracking and tracing process. Using GPS, the data for location will be stored in the private blockchain. This data will be stored from time to time. There are also sensors for detecting any leakage or measuring the weight. If there is a situation where leakage takes place, then IoT sensors will instantly update the information in blockchain and smart contract. Any participant can audit the blockchain later for tracking if there was an issue at the time of transporting or at any phase of the product cycle. RFID enables the receiver to authenticate the purchase. Moreover, blockchain technology will provide the complete security to the product information. The actors can also choose to differentiate among the information to open for all or hide. Eventually, the actors will have the advantage of keeping internal information secured and required information to be disclosed. As shown in Figure 2, smart contract will be responsible for shipment authentication and transaction process.

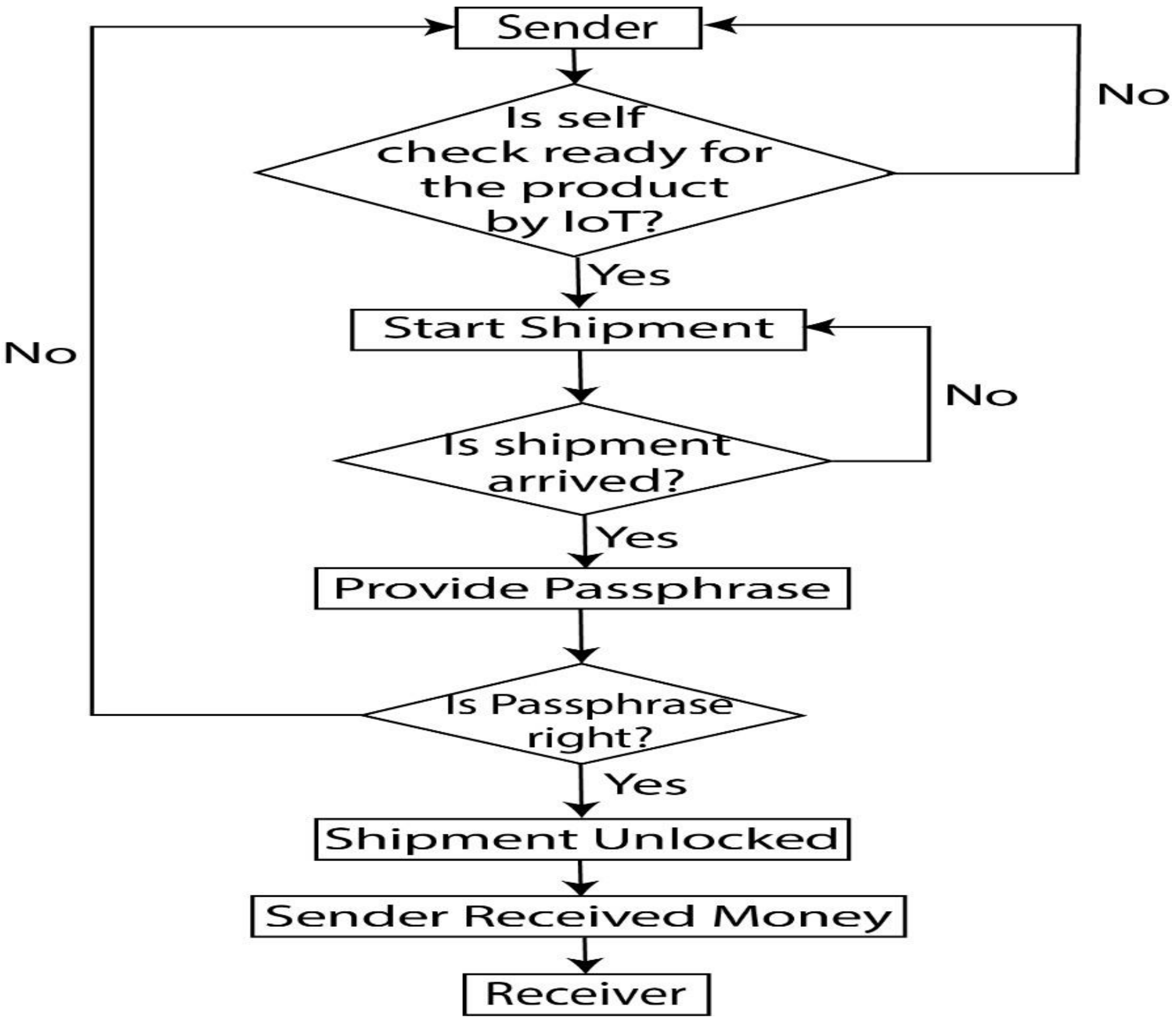

**Figure 2:** Flow Chart for Smart Contract Authentication

### 6.2 Temperature Detection:

One of the most important factors of the distribution of oil is temperature. The container used for transporting oil needs to maintain a certain temperature. Again, there are different types of oil for which different temperature is suitable. Such a concept is used in cold chain management [21]. It is often used in pharmaceutical [22] or food supply chain. But it equally important in oil supply chain. The temperature and humidity sensor will keep track of the environment surrounding the oil. Moreover, it will ensure to cover up any disturbance.

### 6.3 Global and Local Supply Chain

Our framework is proposed for both global and local use case of oil supply chain. It is a world-wide system bringing a lot of entities and participants at one place. Blockchain and smart contract have the capability to cover this huge field of supply chain. Moreover, the transaction process of blockchain using cryptocurrencies e.g., bitcoin also ensures the vividness in economy.

## 7. Implementation and Testing:

In this section, we will discuss the system design and show the output upon testing on the preferred environment.

### 7.1 Smart Contract Architecture:

We aim to design the smart contracts using the Ethereum platform and programming with solidity language. Our smart contract is a basic structure of a complete system. That being said, any user can modify these smart contracts as their needs. The two smart contracts are CheckProgress and OilDistribution. The first contract, CheckProgress is intended to monitor the product's information and keep track of it. It will hold the initial optimum value for package information and also tack the violation. The contract will save this information automatically in the blockchain with the help of IoT devices. The OilDistribution contract will check the authenticity of the actors. The initial pricing and quantity will be stored by it. Table 1 represents the attributes, events, modifiers and functions that are used to implement the smart contract CheckProgress.

**Table 1:** CheckProgress Smart Contract

| **CheckProgress** | | |
|---|---|---|
| Attributes | Events & Modifiers | Functions |
| TempStage: enum<br>HumidityStage: enum<br>PressureStage: enum<br>ViolationType: enum<br>myAddress: address<br>dataAddress: address<br>accurateHum: uint<br>amount: uint<br>totalPrice: uint<br>accurateTemp: uint<br>accuratePress: uint<br>oilID: string<br>oilName: string | oilAdded: event<br>TemperatureViolation: event<br>HumidityViolation: event<br>PressureViolation: event<br>Self (): modifier<br>Checker (): modifier | EnterOil (): Self<br>CheckTemperature ():<br>Checker<br>CheckHumidity (): Checker<br>CheckPressure (): Checker<br>OccuredViolation (): Checker |

Table 2 represents the attributes, events, modifiers and functions that are used to implement the smart contract OilDistribution.

**Table 2**: OilDistribution Smart Contract

| **OilDistribution** | | |
|---|---|---|
| Attributes | Events & Modifiers | Functions |
| CurrentTrace: enum<br>myAddress: address<br>driller_address: address<br>factory_address: address<br>storage_address: address<br>pump_address: address<br>accurateHum: uint<br>drilling_date: uint<br>factory_dist_start_date: uint<br>refiner_start_date: uint<br>pump_start_date: uint<br>drill_price: uint<br>factory_price: uint<br>storage_price: uint<br>pump_price: uint<br>driller_sold_amount: uint<br>factory_sold_amount: uint<br>storage_sold_amount: uint<br>pump_sold_amount: uint<br>oilID: string<br>oilName: string | InitiateDist: event<br>FactoryDistribution: event<br>StorageWholesell: event<br>PumpOilSold: event<br>onlyOwner (): modifier<br>onlyDriller (): modifier<br>onlyFactory (): modifier<br>onlyStorage (): modifier | readyToFactory ():<br>onlyDriller<br>readyToStorage ():<br>onlyFactory<br>oilInOilStorage ():<br>onlyStorage<br>pumpSoldOil (): publlic |

### 7.2 Monitoring Algorithm

Smart contract of each product will follow a monitoring system to track if the product maintains the optimum essential characteristics or violates. The algorithm for this monitoring system is as follows in Table 3:

**Table 3:** Algorithm for monitoring information

| |
|---|
| **Algorithm 01:** Monitoring Algorithm |
| **Output:** Compare input value and accurate value. |
| **Input:** Value received by IoT devices. |
| **if** ( value > accurateValue )<br>    **then** - call occuredViolation( ) method with violation type & stage.<br>    **then** – return a string mentioning high value violation.<br>**else if** ( value < accurateValue )<br>    **then** – call occuredViolation ( ) method with violation type & stage.<br>    **then** – return a string mentioning low value violation.<br>**else**<br>    **then** - call occuredViolation ( ) method with violation type & stage.<br>    **then** – return a string mentioning accurate value violation.<br>**end** |

The data collected by the monitoring algorithm will then be checked for violation. Each of the function for pressure, temperature and humidity will compare the input value with accurate value that was set initially. Finally, these functions will emit the corresponding events to the blockchain to store any kind of violation in the supply chain. The algorithm for OccuredViolation is as follows in Table 4:

**Table 4:** Algorithm for Occurred Violation

| |
|---|
| **Algorithm 02:** Occurred Violation Algorithm |
| **Output:** Triggers a violation event |
| **Input:** ViolationType and ViolationStage |
| **if** violation type == Pressure **then**<br>    **if** stage == 2 **then**<br>        make the enum pressure stage high;<br>        make the enum violation type pressure;<br>        **emit** Pressure Violation event output is Higher Pressure;<br>    **else if** stage == 1 **then**<br>        make the enum pressure stage low;<br>        make the enum violation type pressure;<br>        **emit** Pressure Violation event output is Lower Pressure;<br>    **else if** stage == 0 **then**<br>        make the enum pressure stage accurate;<br>        make the enum violation type none;<br>        **emit** Pressure Violation event output is Accurate Pressure;<br>    **else**<br>        Continue Process;<br>    **end**<br>**else if** violation type == Temperature **then**<br>    similar to pressure;<br>**else if** violation type == Humidity **then**<br>    similar to pressure;<br>**else**<br>    Continue Process;<br>**end** |

To store information of distribution for certain actors, will follow the third algorithm. It takes an actor to initiate distribution with the relevant information. Once an actor initiates the distribution, the method emits the information to the blockchain that the distribution has started and continue further in the supply chain. All the actors can use the algorithm as their personal needs at the time of initiating. The algorithm is as follows in Table 5:

**Table 5**: Algorithm of Oil Distribution

| |
|---|
| **Algorithm 03:** Oil Distribution Algorithm |
| **Output:** Initialization of Actors in Oil Distribution Process |
| **if** caller == actor **then**<br>make the enum PresentTrace to Actor;<br>initialize the time with present timestamp;<br>store id, name, price, quantity in the corresponding variables;<br>**emit** inititateDistribution with actor address and message;<br>**else**<br>continue Process;<br>**end** |

**7.3 Testing and Output**

For the testing purpose, we use Remix IDE. Remix is an online IDE that enables developers to use Ethereum wallets with dummy cryptocurrency. We write the program with the help of solidity language. Another benefit of using Remix is it can compile and deploy smart contract in the blockchain like we have shown in our framework.

The EnterOil() function will hold the information that will be initialized by the smart contract creator as a benchmark. Optimum values like temperature, humidity and pressure will be set here. The output is shown in Figure 3.

hash 0xcd8c5dc9015aacebfd439bf4c94cbb60e08ee3ffe94918bfd708889a131e8e7b

input 0xd84...00000

decoded input { "string name": "Petrol", "string oil_id": "101", "uint256 amt": { "type": "BigNumber", "hex": "0x0a" }, "uint256 price": { "type": "BigNumber", "hex": "0x64" }, "uint256 actualTemp": { "type": "BigNumber", "hex": "0x16" }, "uint256 actualHum": { "type": "BigNumber", "hex": "0x0a" }, "uint256 actualPress": { "type": "BigNumber", "hex": "0x08" } }

**Figure 3:** Enter Oil Output

Then the smart contract will use the values from IoT devices to check with the initialized parameter and show message if any violation happens. The output for pressure is shown in Figure 4.

decoded input { "uint256 value": { "type": "BigNumber", "hex": "0x06" } }

decoded output { "0": "string: Current Pressure is very LOW" }

logs [ { "from": "0x9D7f74d0C41E726EC95884E0e97Fa6129e3b5E99", "topic": "0x14e2bd61e882fc303931e2c975068a4a83834bf571bdf5a403daeffd3d85aea3", "event": "PressureViolation", "args": { "0": "0x5B38Da6a701c568545dCfcB03FcB875f56beddC4", "1": "Lower Pressure", "addr": "0x5B38Da6a701c568545dCfcB03FcB875f56beddC4", "msg": "Lower Pressure" } } ]

**Figure 4**: Check Pressure Output

The output for OilDistribution smart contract is shown in Figure 5. It will show the product ready message for each actor when initiates the smart contract.

logs [ { "from": "0xBBa767f31960394B6c57705A5e1F0B2Aa97f0Ce8", "topic": "0xf3125486c1c86e2b86ddaa332a55424a09dc2326ed15e8b0b76a053b19131a92", "event": "InitiateDist", "args": { "0": "0xAb8483F64d9C6d1EcF9b849Ae677dD3315835cb2", "1": "Crude Oil is Ready to go to the Factory.", "ad": "0xAb8483F64d9C6d1EcF9b849Ae677dD3315835cb2", "masg": "Crude Oil is Ready to go to the Factory." } } ]

**Figure 5**: Ready to Factory Output

## 8. Analysis

In this section, we analyze our output and function execution with respect to efficiency and eligibility in industrial application.

### 8.1 Cost Estimation

The functions that are called in an Ethereum smart contract use gas as the unit for cost estimation. With the type and number operations and complexity varies the amount of gas and so is the cost. The client usually make the gas limit high for a balance among them and miners. A higher gas price increases the probability for a block to be mined because it also increases the reward for mining. Moreover, the smart contract throws an exception if the gas cost becomes higher than the gas limit. The gas limit is set following a balance in speed and cost [23]. In this paper, we try to design our smart contracts with higher gas prices because it ensures an increase in profit for miners and so the blocks are more likely to be picked. Here we use Ethereum Gas Station (EGS) to convert the gas into US dollar as a fiat currency [24]. Gas cost is shown in Table 6.

In the following table, the execution cost is shown for eah of the functions that is used in smart contract. The execution cost is given as gas cost. It is used in Remix IDE for smart contract. This simulates the cost structure of real life implementeation. The system provides some common values for gas cost based on the clients' choice of gas prices for slow, average, fast and fastest execution. As per April 14, 2021, the gas prices is predicted for slow, average, fast and fastest execution are 82, 83, 125 and 147 Gwei per gas. The cost of each function is less than $20.40 for slow or cheap execution, less than $20.65 for average execution, less than $31.10 fast execution and less than $36.57 for fastest execution. Considering the conversion rate of USD according to April 2021, we can say, the cost is justifiable for the presented solution.

**Table 6:** Gas cost for different tasks

| Function Name | Execution Cost | Transaction Cost | Slow Execution (USD) | Avg. Execution (USD) | Fast Execution (USD) | Fastest Execution (USD) |
|---|---|---|---|---|---|---|
| EnterOil | 11408 | 35368 | 2.14323 | 2.16935 | 3.26697 | 3.84201 |
| CheckPressure | 29915 | 51379 | 5.61982 | 5.68832 | 8.56697 | 10.07467 |
| CheckTemperature | 13683 | 35147 | 2.5705 | 2.60189 | 3.91853 | 4.60812 |
| CheckHumidity | 14825 | 36289 | 2.78517 | 2.81908 | 4.24545 | 4.99278 |
| OccuredViolation | 11715 | 33371 | 2.20073 | 2.22754 | 3.35494 | 3.94533 |
| readyToFactory | 108601 | 131857 | 20.40204 | 20.65084 | 31.10055 | 36.57421 |
| ReadyToStorage | 85051 | 106707 | 15.97789 | 16.17263 | 24.35654 | 28.64323 |
| oilInOilStorage | 84865 | 106721 | 15.94284 | 16.13735 | 24.30316 | 28.58068 |
| pumpSoldOil | 68923 | 90579 | 15.94284 | 16.13735 | 24.30316 | 28.58068 |

### 8.2 Advantages

The proposed framework aims to solve different issues of traditional supply chain providing a greater hand on overall system. Followings are the main advantages that the industry can get from the architecture:

- Data Security: Blockchain is the place where all the data is strongly secured. A lot of heterogenous data will be gathered to store in the blockchain from preprocessing of drilling crude oil to selling to the customer. These data will never be changed and almost impossible to tamper.
- Real-Time Data Update: Smart contracts check the value and save it into the blockchain. The system checks these values with different functions automatically. The compatibility between blockchain and smart contracts enable the system to update the information in real-time.
- Customer Access: Two types of blockchain provide everyone the data at any time if needed. The permissioned and hybrid type of blockchain ensures that no unauthorized customer can access.
- Eliminates Third-Party: The use of smart contracts removes the possibility of intermediators to control the oil price making the international market volatile.
- Product Authenticity: Oil id, oil name, date and time are stored in the blockchain. So, last consumer can check the authenticity of the oil.

Hence, the automatization of oil supply chain using our framework can open a new window for oil industry.

**8.3 Challenges and Discussion**

The proposed framework can mitigate the problems of the traditional oil supply chain. However, there are some challenges too.

Firstly, there will be a massive load of data from IoT devices. There will be multiple packages from different companies all around the world. Monitoring each of these packages or transactions is itself a hard task. Our proposed framework can ease the way but the scalability of blockchain is still an issue. We try to increase the latency by avoiding public blockchain. But consortium and private blockchain still a little back from a fast update.

Secondly, the sustainability of the system is a big challenge. For a global supply chain, it can be hard to solve an IoT-based problem instantly. Delay in the update process can eventually cancel the smart contract. Again, lack of backup can result in data loss.

Thirdly, the management of this huge data in blockchain increases the time and power needed for mining. Blockchain holds all the previous information in each block. It is a special benefit of blockchain but at the same time it hampers productivity. Our framework uses a consortium blockchain that uses Byzantine Fault Tolerance (BFT) [25] as its consensus protocol. Following the BFT, a set of elected devices are responsible for authenticity of each block [26]. . It usually utilizes fewer resources than another type of blockchain.

Fourthly, the maintenance of a global supply chain is a tedious task. The area of oil supply chain is vast and complex. The use of blockchain and smart contracts can lessen the tasks. However, to increase efficiency at a global scale, the maintenance procedure must be more robust.

Finally, the interaction among different types of blockchain and IoT devices is still an issue. The wireless interaction in the system arises an optimization issue as well as a security issue. So, there is a need to enhance the ability of IoT devices.

## 9. Conclusion and Future Research

In this paper, we have tried to contribute to modifying and modernizing the oil supply chain for better and smarter. Our framework tends to provide complete transparency providing immutable product information. It ensures prevention from other counterfeiting entities to enter the system. Moreover, the use of smart contract reduces the hard paper work. Usually, there needs much paperwork in the traditional oil supply chain. Blockchain makes these smart contracts immutable keeping the information as proof. However, the massive amount of data can affect the latency. Off-chain [27] transactions can be useful in terms of storage and computational expenses. There are other approaches to increase the throughput by implementing decentralized databases like BigchainDB [28] and HBasechainDB [29]. Besides, in case of large scale implementation the sensor devices can be added to the frameworkand more actors can also be added. Since the amount of data will be increased we can use off-chain architecture to hold the original data and the proof of existence can be stored on blockchian itself. This is can be taken into account as a future research direction of this study

Our framework contributes mostly to the monitoring and traceability process in the oil supply chain. We have conceptualized a process for end-to-end tracking of products. It also provides all the consumers with product information at any time from anywhere. We have identified the role and function of every actor. We also imply the applications of our framework for different purposes. The structure of smart contracts is also described. They can be modified as required. The detail code is uploaded in the GitHub repository: https://github.com/RifatShovon/SmartOil. With some further development in management and database systems, our framework can be used by any real-world company to construct, adapt and maintain smart oil supply chain management.